\documentclass[prl,superscriptaddress,a4paper,twocolumn]{revtex4-1}
\usepackage{geometry}
\usepackage[english]{babel}
\usepackage[latin1]{inputenc}
\usepackage{hyperref}
\usepackage{amsmath, amssymb, amsfonts, mathrsfs}
\usepackage{graphicx}

\newcommand{\ket}[1]{|#1\rangle}

\begin{document} 

\title{Quantum Experiments and Graphs:\\ Multiparty States as coherent superpositions of Perfect Matchings}

\author{Mario Krenn}
\email{mario.krenn@univie.ac.at}
\affiliation{Vienna Center for Quantum Science \& Technology (VCQ), Faculty of Physics, University of Vienna, Boltzmanngasse 5, 1090 Vienna, Austria.}
\affiliation{Institute for Quantum Optics and Quantum Information (IQOQI), Austrian Academy of Sciences, Boltzmanngasse 3, 1090 Vienna, Austria.}
\author{Xuemei Gu}
\affiliation{Institute for Quantum Optics and Quantum Information (IQOQI), Austrian Academy of Sciences, Boltzmanngasse 3, 1090 Vienna, Austria.}
\affiliation{State Key Laboratory for Novel Software Technology, Nanjing University, 163 Xianlin Avenue, Qixia District, 210023, Nanjing City, China.}
\author{Anton Zeilinger}
\email{anton.zeilinger@univie.ac.at}
\affiliation{Vienna Center for Quantum Science \& Technology (VCQ), Faculty of Physics, University of Vienna, Boltzmanngasse 5, 1090 Vienna, Austria.}
\affiliation{Institute for Quantum Optics and Quantum Information (IQOQI), Austrian Academy of Sciences, Boltzmanngasse 3, 1090 Vienna, Austria.}

\begin{abstract}
We show a surprising link between experimental setups to realize high-dimensional multipartite quantum states and Graph Theory. In these setups, the paths of photons are identified such that the photon-source information is never created. We find that each of these setups corresponds to an undirected graph, and every undirected graph corresponds to an experimental setup. Every term in the emerging quantum superposition corresponds to a perfect matching in the graph. Calculating the final quantum state is in the complexity class \textsc{\#P-complete}, thus cannot be done efficiently. To strengthen the link further, theorems from Graph Theory -- such as Hall's marriage problem -- are rephrased in the language of pair creation in quantum experiments. We show explicitly how this link allows to answer questions about quantum experiments (such as which classes of entangled states can be created) with graph theoretical methods, and potentially simulate properties of Graphs and Networks with quantum experiments (such as critical exponents and phase transitions).
\end{abstract}
\date{\today}
\maketitle 

When a pair of photons is created, and one cannot -- even in principle -- determine what its origin is, the resulting quantum state is a coherent superposition of all possibilities \cite{wang1991induced, zou1991induced}. This phenomenon has found a manifold of applications such as in spectroscopy \cite{kalashnikov2016infrared}, in quantum imaging \cite{lemos2014quantum}, for the investigation of complementarity \cite{heuer2015induced}, in superconducting cavities \cite{lahteenmaki2016coherence} and for investigating quantum correlations \cite{hochrainer2017quantifying}. By exploiting these ideas, the creation of a large number of high-dimensional multipartite entangled states has been proposed recently \cite{krenn2017entanglement} (inspired by computer-designed quantum experiments \cite{krenn2016automated}).

\begin{figure}[ht]
\includegraphics[width=0.5 \textwidth]{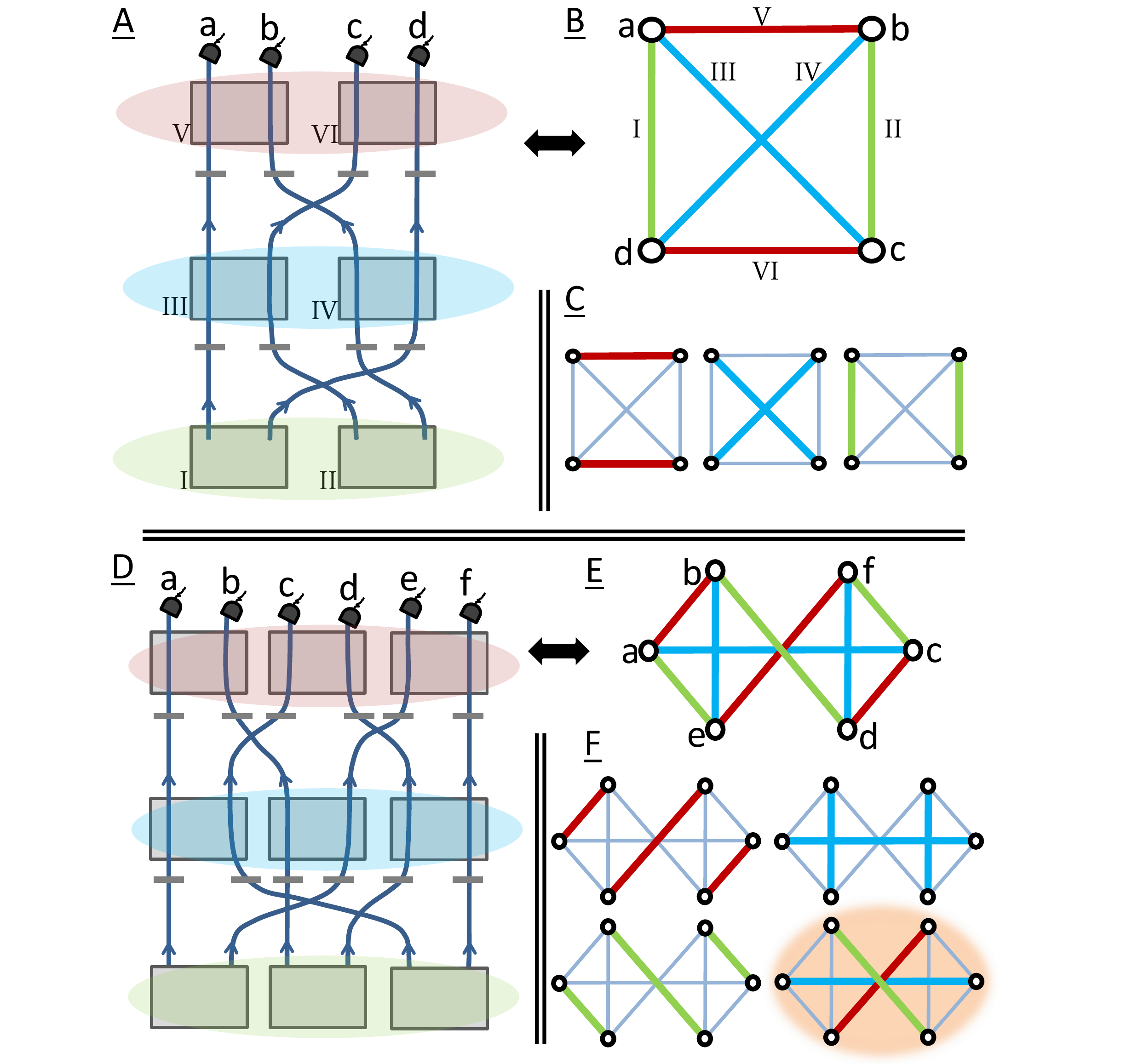}
\caption{\textbf{A}: An optical setup which can create a 3-dimensional 4-photon GHZ-state with the method of \textit{Entanglement by Path Identity} \cite{krenn2017entanglement}. It consists of three layers of crystals, in between there are variable mode- and phase-shifters (depicted in grey). \textbf{B}: The corresponding graph with four vertices (one for each path), six edges (one for each crystal). Every layer of crystals leads to a four-fold coincidence count.  \textbf{C}: These corresponds to three disjoint \textit{perfect matching} or \textit{1-factors} in the graph. \textbf{D}: An optical setup for creating 3-dimensional entanglement with 6 photons. \textbf{E}: The corresponding graph \textbf{F}: It has four perfect matchings, thus the corresponding quantum state has four terms. One terms comes from each of the three layers (the GHZ terms), and one additional term comes from different layers (the \textit{Maverick}-term, with orange background). For that reason, the resulting quantum state has not the form of a GHZ state. In the Appendix, we show how to construct the experimental setup from a given graph.}  
\label{fig:3dim4party}
\end{figure}

Here we show that Graph Theory is a very good abstract descriptive tool for such quantum experimental configuration: Every experiment corresponds to an undirected Graph, and every undirected Graph is associated with an experiment. On the one hand, we explicitly show how to translate questions from quantum experiments and answer them with graph theoretical methods. On the other hand, we rephrase theorems in Graph Theory and explain them in terms of quantum experiments. 

An important example for this link is the number of terms in the resulting quantum state for a given quantum experiment. It is the number of perfect matchings that exists in the corresponding graph -- a problem that lies in the complexity class \textsc{\#P-complete} \cite{valiant1979complexity}. Futhermore, the link can be used as a natural implementation for the experimental investigation of quantum random networks \cite{perseguers2009quantum}.

\begin{table}[b]
  \centering
    \begin{tabular}{ | l | l |}
    \hline
    \textbf{Quantum Experiment} & \textbf{Graph Theory}\\ \hline \hline
    Optical Setup with Crystals & undirected Graph $G(V,E)$ \\ \hline
    Crystals & Edges $E$ \\ \hline
    Optical Paths & Vertices $V$ \\ \hline
    n-fold coincidence & perfect matching \\ \hline
    \#(terms in quantum state) & \#(perfect matchings) \\ \hline
    maximal dimension of photon & degree of vertex \\ \hline
    \shortstack{n-photon d-dimensional \\ GHZ state} & \shortstack{n-vertex graph with\\ d disjoint perfect matchings} \\ \hline                 
    \hline
    \end{tabular}
  \caption{The analogies between Quantum Experiments involving multiple crystals and Graph Theory.}
  \label{tab:compare}
\end{table}

\textit{Experiments and Graph} -- The optical setup for creating a 3-dimensional generalization of a 4-photon Greenberger-Horne-Zeilinger state \cite{greenberger1989going, lawrence2014rotational} is shown in Fig. \ref{fig:3dim4party}A \cite{krenn2017entanglement}. The experiment consists of three layers of two down-conversion crystals each. Each crystal can create a pair of photons in the state $\ket{0,0}$, where the mode number could correspond to the orbital angular momentum (OAM) of photons \cite{allen1992orbital, yao2011orbital, krenn2017orbital} or some other (high-dimensional) degree-of-freedom. A laser pumps all of the six crystals coherently, such that two pairs of photons are created in parallel. Four-fold coincidence (i.e. four photons are detected simultaneously in detector $a$, $b$, $c$ and $d$) can only happen if the two photon pairs are created in crystals I and II, or in crystals III and IV or in crystals V and VI. In every other case, there is at least one path without a photon, which is neglected by post-selection. Between each layer, the modes are shifted by $+1$. This example leads to the final state $\ket{\psi}=1/\sqrt{3}\left(\ket{0,0,0,0}+\ket{1,1,1,1}+\ket{2,2,2,2}\right)$.

The corresponding graph is shown in Fig. \ref{fig:3dim4party}B. Every optical path $a$, $b$, $c$, $d$ in the experiment corresponds to a vertex in the graph, every crystal forms an edge between the vertices. A four-fold coincidence count happens if a subset of the edges contain each of the four vertices exactly once. Such a subset is called \textit{perfect matching} of the graph. In the above example, there are three perfect matchings (two green edges, two blue edges and two red edges), thus there are three terms in the quantum state. We can therefore think of our quantum state as a coherent superposition of the perfect matchings in the corresponding graph. The correspondence between quantum optical setups and graph theoretical concepts are listed in Table \ref{tab:compare}.

Now, what will happen when we add more crystals in each layer? As an example, in Fig. \ref{fig:3dim4party}D, three crystals in each layer produce 6 photons, there are three layers which make the photons 3-dimensionally entangled. Surprisingly however, in contrast to the natural generalisation of the 4-photon case in Fig. \ref{fig:3dim4party}A-C (and in contrast to what some of us wrote in \cite{krenn2017entanglement}), the resulting state is not a high-dimensional GHZ state. In contrast to the previous case, there are four perfect matchings, thus the resulting quantum state has four terms (Fig. \ref{fig:3dim4party}F). One perfect matching comes from each of the layers (which are the terms expected for the GHZ state), and one additional arises due to a combination of one crystal from each layer (which we call \textit{Maverick}-term). If the mode shifter between the layers is $+1$ as before, the Maverick term has $\ket{1_a,1_c}$ from the blue layer, $\ket{2_b,2_d}$ from the green layer and $\ket{0_e,0_f}$ from the red layer. This leads to the final state
\begin{align}
\ket{\psi}=\frac{1}{2}\big(&\ket{0,0,0,0,0,0}+\ket{1,1,1,1,1,1} \nonumber \\
&+\ket{2,2,2,2,2,2}+\ket{1,2,1,2,0,0} \big).
\label{eq:3dim6party}
\end{align}
A GHZ state can only appear when all perfect matchings are \textit{disjoint}, meaning that every edge appears only in one perfect matching. Otherwise, additional terms are present in the quantum state. 

\begin{figure*}
\includegraphics[width=\textwidth]{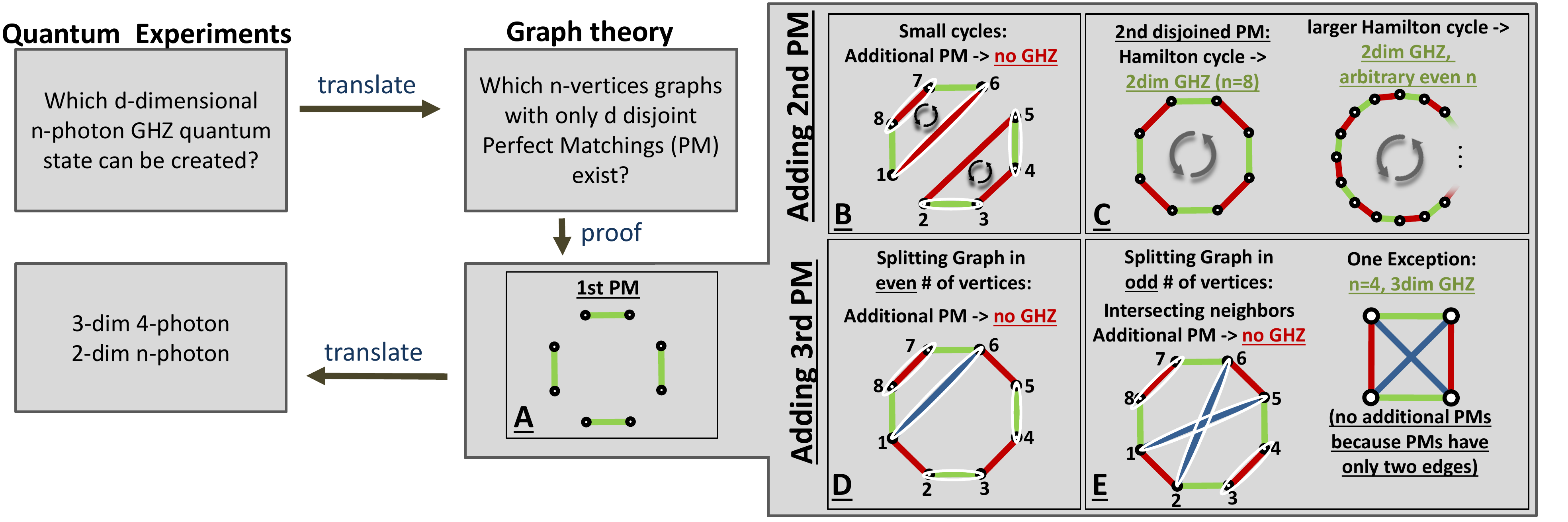}
\caption{Application of the bridge between quantum experiments and graph theory: As a concrete example, we ask which $d$-dimensional $n$-photon GHZ states can be created experimentally with this method. The idea of the proof is to construct a graph starting with $n$ vertices without edges. We try to maximize the number of \textit{disjoint} perfect matchings (PMs) by adding appropriate edges to the graph \cite{267013}. In disjoint PMs, every edge appears only in one perfect matching. The example in the figure is for $n$=8, but the proof works for arbitrary even $n$. \textbf{\underline{STEP I}}: In \textbf{A}, we add the first PM to a set of 8 vertices (green). \textbf{\underline{STEP II}}: In \textbf{B}, we add more edges to construct a second PM (red). Whenever the new PM together with the first (green) PM creates more than one cycle (here: edges 1-6,6-7,7-8,8-1; and 2-3,3-4,4-5,5-2), we immediatly find an additional Maverick PM (indicated with white boundary, edges 1-6,2-3,4-5,7-8). Thus the graph cannot represent a GHZ state (as a GHZ state has only disjoint perfect matchings). The only choice for the second PM is to create together with the first PM \textit{one} cycle which visits every vertex -- a Hamilton cycle, shown in \textbf{C}. Hamilton cycles consist of 2 PMs, therefore correspond to 2-dimensional GHZ states. It can be arbitrarily large, thus there can be arbitrarily large $n$-photon 2-dimensional GHZ states. \textbf{\underline{STEP III}}: Starting with the Hamilton cycle, we try to add a third PM, with blue edges. In \textbf{D} we observe that if the new edge splits the graph into an even number of vertices (upper part: vertices 7,8; lower part: vertices 2,3,4,5), we always find a new Maverick PM. It consists of the new edge (here: 1-6) and edges from the Hamilton cycle (here edges 2-3,4-5,7-8). We learn -- as we require only disjoint perfect matchings -- no edge of a new PM should split the graph into even numbers of vertices (otherwise Maverick PMs appear). Finally in \textbf{E} we try to add edges which split the graph into odd number of vertices. We observe that in every additional PM there are at least two neighboring edges which intersect (neighboring edges start from consecutive vertices; here -- shown in blue -- they start at vertex 1 and vertex 2). This pair always forms a new Maverick PM with additional edges from the Hamilton cycle (here: 1-5,2-6,3-4,7-8). There is one exception for the case of $n$=4: There can be a 3rd disjoint PM, because a Maverick PM needs at least 3 edges (2 blue ones and one from the Hamilton cycle). Therefore, a 4-photon 3-dimensional GHZ state can be created, while for $n>4$, GHZ states can only be created with $d=2$.}
\label{fig:ProofGraph}
\end{figure*}

When the number of layers of crystals is increased to four (with 3 cystals per layer) and modes are shifted by +1 as before (and no phase-shifters are used), there are 8 terms in the resulting quantum state: 4 GHZ-like terms and 4 additional Maverick terms. For five layers, the resulting 6-photon quantum state consists of 15 terms (5 GHZ-like terms and 10 additional Maverick terms), entangled in 5 dimensions (see Appendix). In general, $n$ crystals in one layer produce 2$n$ photons. One can design setups with $d=(2n-1)$ layers, which correspond to a complete graph $K_{2n}$ (in a complete graph, every vertex is connected with every other one exactly once). It produces a state with $\frac{(2n)!}{n! 2^n}$ terms, $(2n-1)$ of them are GHZ-like (see Appendix). By changing the mode shifters and phase shifters between the layers, a vast amount of different quantum states can be created.

Now one could ask what types of GHZ states are possible in general using the experimental scheme above. We show a proof based on Graph Theory which answers that question. For that, we first translate the quantum physics question \textit{Which $d$-dimensional GHZ states can be created?} into the graph theory question \textit{Which undirected graphs exist with $d$ perfect matchings which all are disjoint?}. The proof strategy is to construct a graph with a maximum number of disjoint perfect matchings, starting from $n$ vertices \cite{267013}. The concept and the proof are described in Fig. \ref{fig:ProofGraph}. We find that one can create arbitrarily large 2-dimensional GHZ states, and a 3-dimensional 4-photon GHZ state. In an analogous way, different questions in such quantum experiments can be translated and answered with Graph Theory. 
 
In order to build 3-dimensional GHZ-type experiments with 6 photons (without extra terms), one can use two copies of the 3-dimensional 4-photon GHZ state (presented in Fig. \ref{fig:3dim4party}A), and combine them with a 3-dimensional Bell-state measurement \cite{bennett1993teleporting, sych2009complete}. In the graph this is represented by two graphs that are merged (see Appendix). Many other classes of entangled states, such as two-dimensional W-state \cite{zeilinger1992higher, bourennane2004experimental} or asymmetrically entangled Schmidt-Rank Vector (SRV) \cite{huber2013structure, malik2016multi} can be created by exploiting multigraphs (graphs with more then one edge between two vertices), as shown in the Appendix.

\begin{figure}[ht]
\includegraphics[width=0.5 \textwidth]{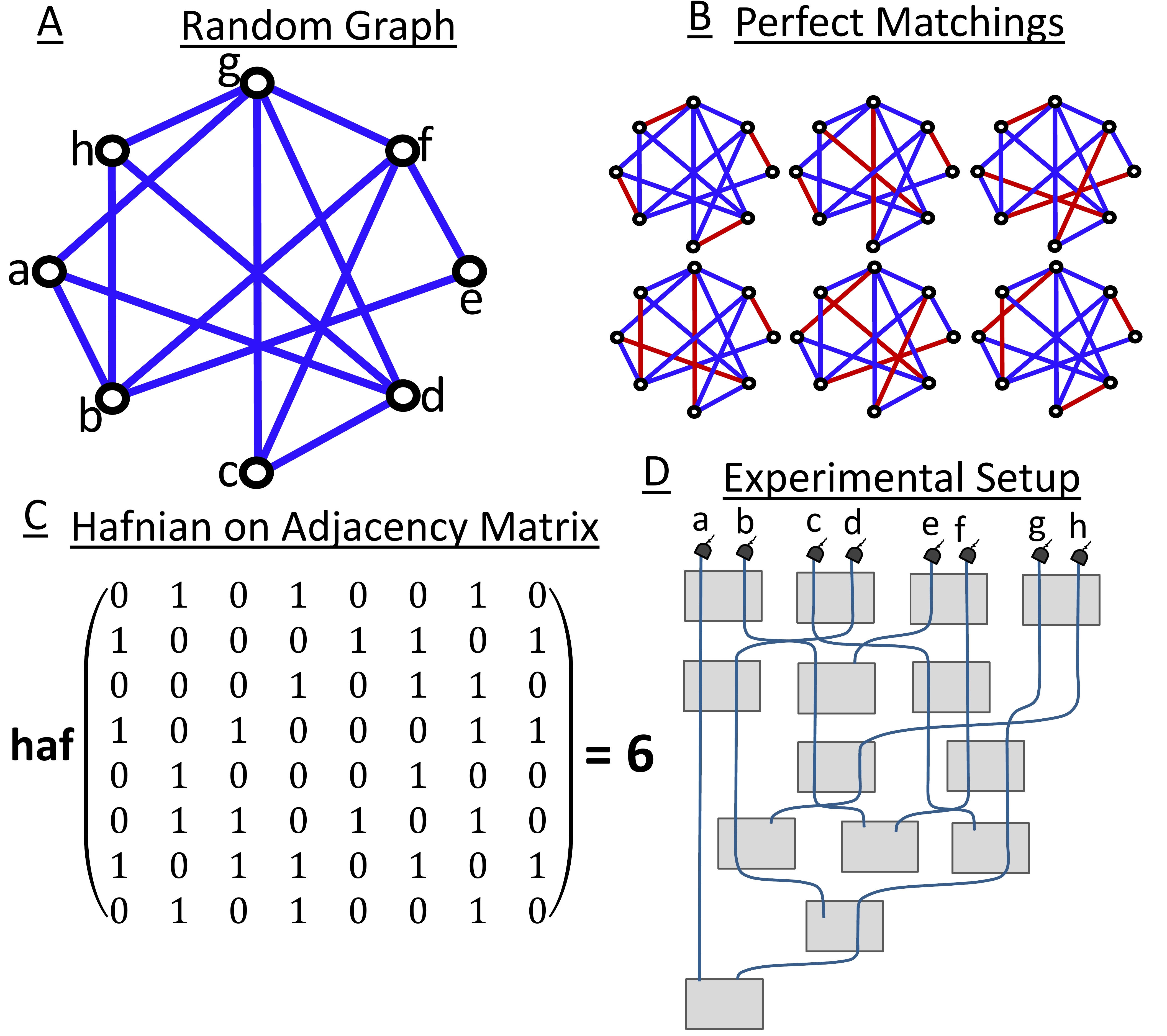}
\caption{Random Graph or Quantum Random Network -- and its connection to Quantum experiments. \textbf{A}: A random graph with 8 vertices and 14 edges. \textbf{B}: The perfect matchings corresponding to the random graph. \textbf{C}: They can be calculated with the matrix function \textit{Hafnian}, which is a generalisation of the \textit{permanent}. Both are very expensive to calculate. \textbf{D}: The corresponding quantum experiment. Each of terms in its quantum state corresponds to a perfect matching in the graph. It can also be seen as a quantum random network, to study network properties in the quantum regime.}  
\label{fig:RandomGraph}
\end{figure}

An important result is that calculating the final quantum state cannot be done efficiently: Counting the number of perfect matchings in a graph (i.e. calculating the number of terms in the resulting quantum state) is in the complexity class \textsc{\#P-complete}. In a bipartite graph, it is equivalent to computing the permanent of the graph's biadjacency matrix \cite{valiant1979complexity} (see Appendix for such an experimental setup). Furthermore, for general graphs, counting the number of perfect matchings corresponds to calculating the \textit{Hafnian} (a generalisation of the permanent) of the graph's adjacency matrix. Even for approximating the Hafnian there is no known deterministic algorithm which runs in polynomial time \cite{bjorklund2012counting, barvinok2017approximating}. An example is given in Fig. \ref{fig:RandomGraph}A for a random graph, its corresponding perfect matching and Hafnian in Fig. \ref{fig:RandomGraph}B-C, and the corresponding quantum setup in Fig. \ref{fig:RandomGraph}D.  

\begin{figure}[ht]
\includegraphics[width=0.5 \textwidth]{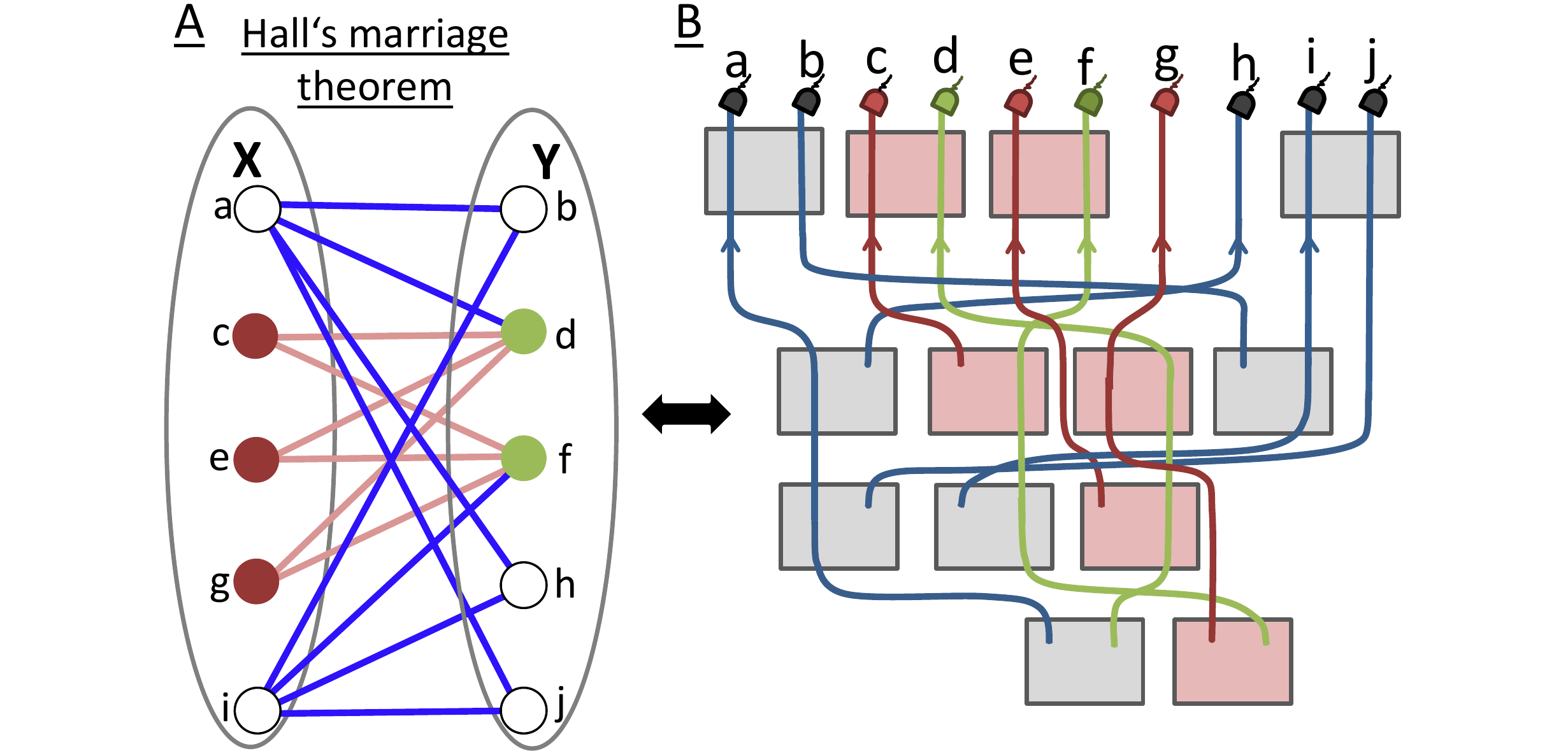}
\caption{A theorem from Graph theory: Hall's marriage theorem \textbf{A}: For a bipartite graph with equal number of elements in $X$ and $Y$, Hall's theorem gives a necessary and sufficient condition for the existence of a perfect matching. That happens when for every subset in $W \in X$, the number of neighbors in $Y$ is larger or equal than $|W|$. In the example graph, the subset of $X$ consisting of the vertices c, e, g (indicated in red) have only two neighbors in $Y$ (d, f -- indicated in green), thus there cannot be a perfect matching. \textbf{B}: For quantum experiments, the analog question is whether there can be 2$n$-fold coincidences, given that $n$ crystals emit photon pairs. When the two photons are distinguishable (which corresponds to a bipartite graph), 2n-folds can only happen when for every subset $W$ of signal photon paths the number of connected idler paths is larger or equal than $|W|$. In the example, the subset of signal photon paths (c, e, g -- depicted in red) has only two corresponding idler paths (d, f -- depicted in green), thus there cannot be a 10-fold coincidence count.}  
\label{fig:GraphTheoryTheorems}
\end{figure}
While the information about the number of terms is encoded in every $n$-photon quantum state emerging from the setup, the question is how one can obtain this information (or approximate it) efficiently. Measurements in the computation basis are not sufficient, otherwise it could be calculated classically as well. One direction would be to investigate \textit{frustrated generation} of multiple qubits \cite{herzog1994frustrated} (for instance, by using phase shifters instead of mode shifters between each crystal), or by analysing multi-photon high-dimensional entanglement detections \cite{huber2010detection}. A detailed investigation of the link between the outcome of such experiments and complexity classes would be valuable, but is outside the scope of this article. 

As it is possible to generate experimental setups for arbitrary undirected graphs, the presented scheme is also a natural and inexpensive implementation of quantum (random) networks (see Fig. \ref{fig:RandomGraph}). This could be used to experimentally investigate entanglement percolation \cite{acin2006entanglement, cuquet2009entanglement, perseguers2010multipartite} and critical exponents which lead to phase transitions in quantum random networks \cite{perseguers2009quantum}. As an example, it has been shown that for large quantum networks with $N$ nodes, every quantum subgraph can be extracted with local operations and classical communication (LOCC) if the edges are connected with a probability $p \geq N^{-2}$ \cite{perseguers2009quantum}. In close analogy to the experimental schemes here, $N$ is the number of output paths of photons, and $p$ corresponds to the probability for a down-conversion event in a single crystal. The quantum state for the edge between vertices $a$ and $b$, with mode number $\ell$ can be written as 
\begin{align}
\ket{\psi_{a,b}}&=\Big(1+p\left(\hat{a}_{a,\ell}^{\dagger} \hat{a}_{b,\ell}^{\dagger} - \hat{a}_{a,\ell} \hat{a}_{b,\ell}\right)+ \nonumber\\
&+ \frac{p^2}{2} \left(\hat{a}_{a,\ell}^{\dagger} \hat{a}_{b,\ell}^{\dagger} - \hat{a}_{a,\ell} \hat{a}_{b,\ell}\right)^2 + ...\Big)\ket{0}.
\label{SPDC}
\end{align}
where $p$ is the SPDC probability. The complete quantum (random) network is a combination of all crystals being pumped coherently, which is a tensor product over all existing edges in the form of
\begin{align}
\ket{\psi_{\textnormal{network}}}=\bigotimes_{e(i,j)\in E}\ket{\psi_{i,j}}
\label{SPDC}
\end{align} 
where $i$ and $j$ are the vertices which are connected by the edge $e \in E$. 

Finally, to strengthen the link between quantum experiments and graph theory, we show that theorems from Graph theory can be translated and reinterpreted in the realm of quantum experiments. In Fig. \ref{fig:GraphTheoryTheorems}A and B, we show \textit{Hall's marriage theorem}, which gives a necessary and sufficient condition in a bipartite graph for the existence of at least one perfect matching \cite{hall1935representatives}. A generalisation to general graphs, \textit{Tutte's theorem} \cite{tutte1947factorization, akiyama2011factors}, is shown in the Appendix. Both Graph theory theorems can be understood in the language of quantum experiments.

To conclude, we have shown a strong link between quantum experiments and Graph Theory. It allows to systematically analyse the emerging quantum states with methods from graph theory. The new link immediatly opens up many new directions for future research. For example, the analysation of the number of \textit{maximal matchings} and \textit{matchings} in a graph (called \textit{Hosoya index} and often used in chemistry \cite{hosoya2002topological, jerrum1987two}) in the contect of quantum experiment.

A detailed investigation of links between these experiments and computation complexity classes, in particular the relation to computation complexity with linear optics would be interesting \cite{aaronson2011computational, aaronson2005quantum, hamilton2016gaussian}. 

Furthermore it would be interesting how the merging of graphs can be generalized with non-destructive measurements \cite{wang2015quantum}, whether it leads to larger classes of accessible states and how that can be described in the Graph theoretical framework.

The generalisation to other graph theoretical methods would be interesting, such as weighted graphs (which could correspond to variable down-conversion rates via modulating the laser power), hypergraphs (which would correspond to creation of tuples of photons, for instance via cascaded down-conversion \cite{hubel2010direct,hamel2014direct}) or 2-Factoriations (or general $n$-Factorizations, which would lead to $n$ photons in one single arm). 

Experimental implementations could not only create a vast array of well-defined quantum states, but could also investigate striking properties of quantum random networks in the laboratory.

Finally, we suggest that recent developments of integrated optics implementations of quantum experiments, where the photons are generated on a photonic chip \cite{silverstone2014chip, jin2014chip, krapick2016chip}, could be particularly useful to realize setups of the type proposed here.

\section*{Acknowledgements}
The authors thank Manuel Erhard, Armin Hochrainer and Johannes Handsteiner for useful discussions and valuable comments on the manuscript. X.G. thanks Lijun Chen for support. This work was supported by the Austrian Academy of Sciences (ÖAW), by the European Research Council (SIQS Grant No. 600645 EU-FP7-ICT) and the Austrian Science Fund (FWF) with SFB F40 (FOQUS). XG acknowledges support from the Major Program of National Natural Science Foundation of China (No. 11690030, 11690032), the National Natural Science Foundation of China (No.61272418).
\bibliographystyle{unsrt}
\bibliography{refs}

\newpage
\widetext

\section{Appendix}

\subsection{Appendix I. Examples of Path Identity}
\begin{figure}[b]
\includegraphics[width=0.75 \textwidth]{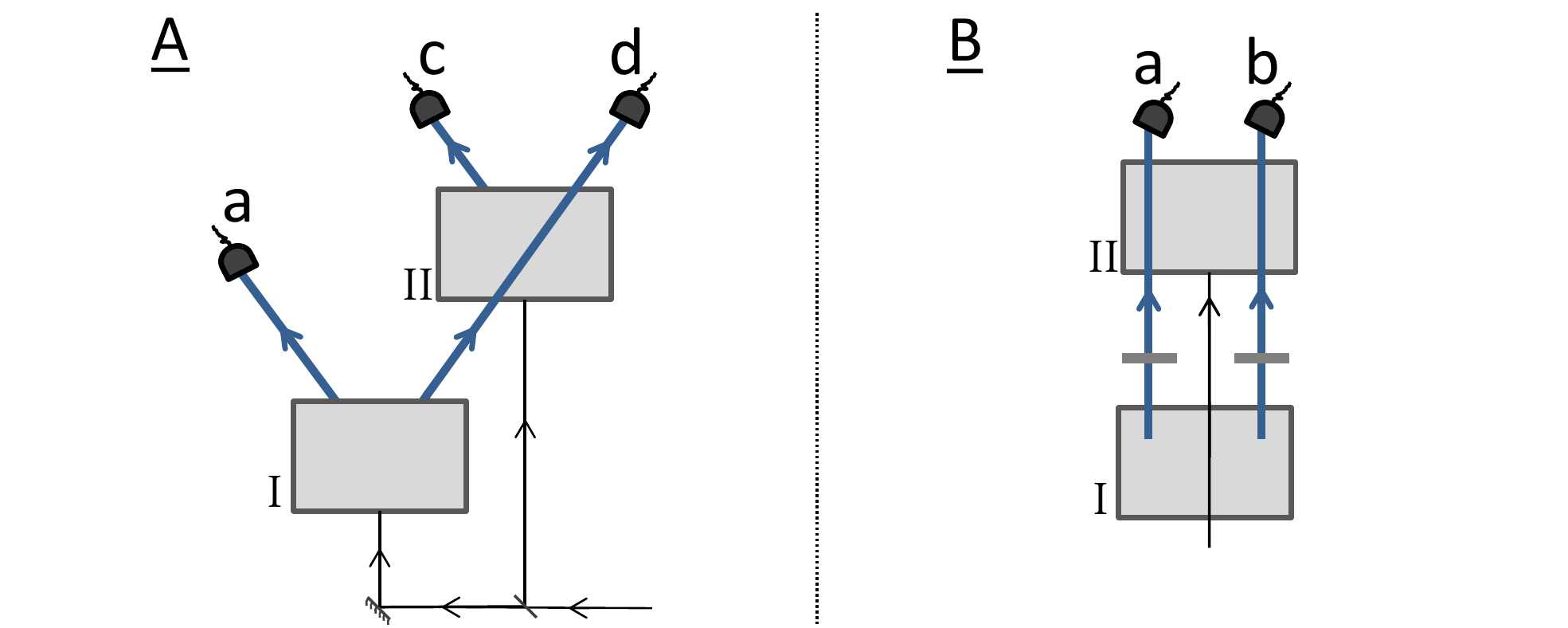}
\caption{\textbf{A}: The experiment introduced in \cite{wang1991induced} consists of two crystals, pumped by a laser (depicted in black) which create one pair of photons (either in crystal I or in crystal II), and one of the paths is overlapped. If the two possibilities are prepared such that one cannot distinguish in which crystal the photons have been created, the final state consists of a photon in $d$ and a coherent superposition of the second photon being in $a$ or in $c$. \textbf{B}: In this experiment, both arms are overlapped. If the grey elements between the two crystals are phase-shifters, the two crystals can either constructively or destructively interfere, leading to larger or smaller numbers of photons in the output $a$ and $b$ \cite{herzog1994frustrated}. If the grey elements are mode-shifters, one creates an entangled state \cite{kwiat1999ultrabright}. In can be chosen by the experimentalist whether the photons emerge colinear or at an angle from the crystal. For simplicity, the laser is not drawn anymore in the following examples.}  
\label{fig:intro}
\end{figure}

Twenty-five years ago, Wang, Zou and Mandel (originally suggested by Zhe-Yu Ou) have demonstrated a remarkable idea: They coherently overlapped one of the output modes from each crystal ($\ket{b}=\ket{d}$ in Fig. \ref{fig:intro}A), such that the \textit{which-crystal information} for the photon in $d$ never exists in the first place. That leads to $\ket{\psi}=1/\sqrt{2} \left(\ket{a}+\ket{c}\right)\ket{d}$, where one photon is in $d$ and the second photon is in a coherent superposition of being in $a$ and in $c$.
When both output modes from the two crystals are overlapped such that the paths of the photons are identical (Fig. \ref{fig:intro}B). By adding phases between the two crystals, one obtains $\ket{\psi}=(\ket{a,b}+\exp(i\phi)\ket{a,b})=(1+\exp(i\phi))\ket{a,b}$, which means that by changing the phase $\phi$, one can enhance or suppress the creation of photons -- a phenomenon denoted as \textit{frustrated} generation of photon pairs \cite{herzog1994frustrated}. If instead of phase shifters one would add mode shifters between the crystals (for instance, the crystal produces two horizontal polarized photons, and the mode-shifter changes horizontal to vertical), one creates an entangled two-photon state $\ket{\psi}=1/\sqrt{2}(\ket{H_a,H_b}+\ket{V_a,V_b})$ \cite{kwiat1999ultrabright}. 

\newpage
\subsection{Appendix II. Multi-Graphs for different structured entangled states}
\begin{figure}[ht]
\includegraphics[width=0.7 \textwidth]{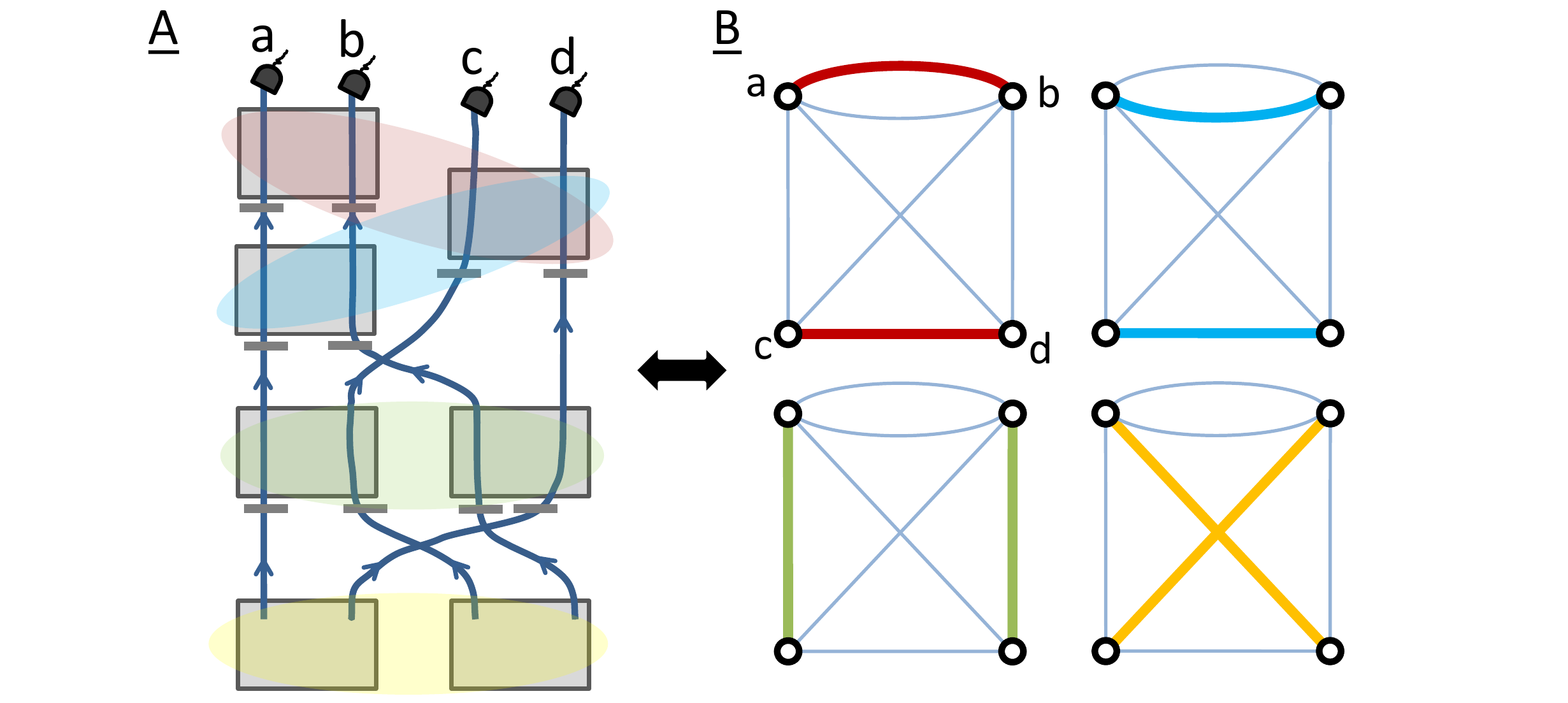}
\caption{\textbf{A}: An optical setup, where two crystals emit into the same path, can be used to realize many entangled states, such as the W-state $\ket{\psi}=1/2\left(\ket{0,0,0,1}+\ket{0,0,1,0}+\ket{0,1,0,0}+\ket{1,0,0,0}\right)$ or a high-dimensional asymmetrically entangled state $\ket{\psi}=1/2 \ket{0}\left(\ket{0,0,0}+\ket{1,0,1}+\ket{2,1,0}+\ket{3,1,1}\right)$, where one photon acts as trigger. By changing the mode- and phase-shifters between the crystals, one arrives at different states. \textbf{B}: Such experiments can be consistently described with multiple edges that are incident to the same two vertices. By looking at the perfect matchings, it is easy to understand what modes the individual crystals have to produce to obtain the desired state (for example, shown in \cite{krenn2017entanglement}).}  
\label{fig:Asymmetric}
\end{figure}

The examples in Fig. 1 and Fig. 2 in the main text have aimed at producing GHZ states. One can use multi-graphs (which have two edges between two vertices) to create a vast array of different entangled states, such as the W-state or high-dimensional asymmetrically entangled states -- as shown in Fig. \ref{fig:Asymmetric}.

\subsection{Appendix III. High-dimensional Multipartite Entanglement Swapping for State merging}
\begin{figure}[ht]
\includegraphics[width=0.7 \textwidth]{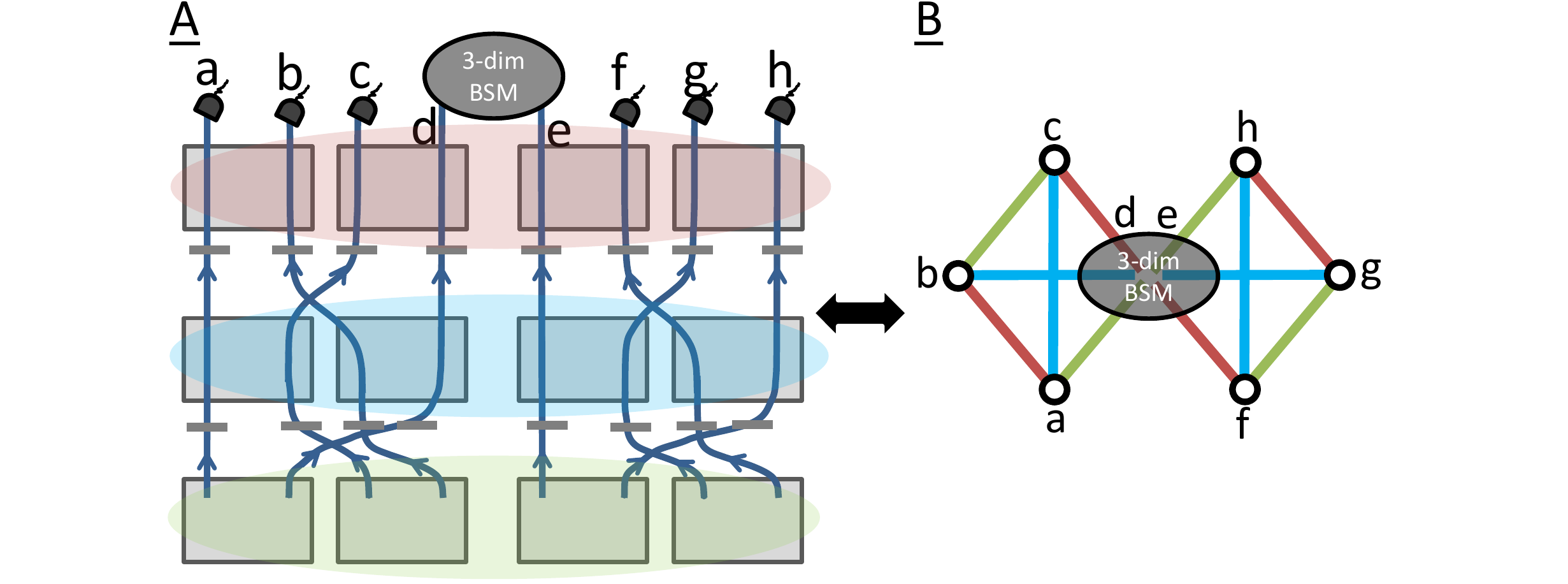}
\caption{\textbf{A}: Two experiments which each create a 3-dimensional 4-photon entangled GHZ state can be combined with a 3-dimensional Bell-State measurement. \textbf{B}: In the corresponding graph, the vertices $d$ and $e$ are merged. Merging the two graphs can be understood as a generalized multi-photon high-dimensional entanglement swapping.}  
\label{fig:EntSwapp}
\end{figure}

The merging of two 4-photon 3-dimensional entangled states with a 3-dimensional Bell-state measurement (as shown in Fig. \ref{fig:EntSwapp}A and B) leads to a 6-photon 3-dimensional entangled state:
\begin{align}
\ket{\psi}=\frac{1}{\sqrt{3}}\big(&\ket{0,0,0,0,0,0}+\ket{1,1,1,1,1,1}+\ket{2,2,2,2,2,2} \big),
\label{eq:3dim6partyGHZ}
\end{align}
which is a 6-photon, 3-dimensional GHZ state. This can be generalized to multi-photon 3-dimensional GHZ states with more copies chained together. The operation is a generalisation of entanglement swapping \cite{zukowski1993event, pan1998experimental} to multi-photonic systems \cite{bose1998multiparticle} with more than two dimensions. 
        
\newpage

\subsection{Appendix VI. Entanglement of 6 photons in 5 dimensions - complete graph $K_{6}$}
\begin{figure*}[!h]
\includegraphics[width=0.9 \textwidth]{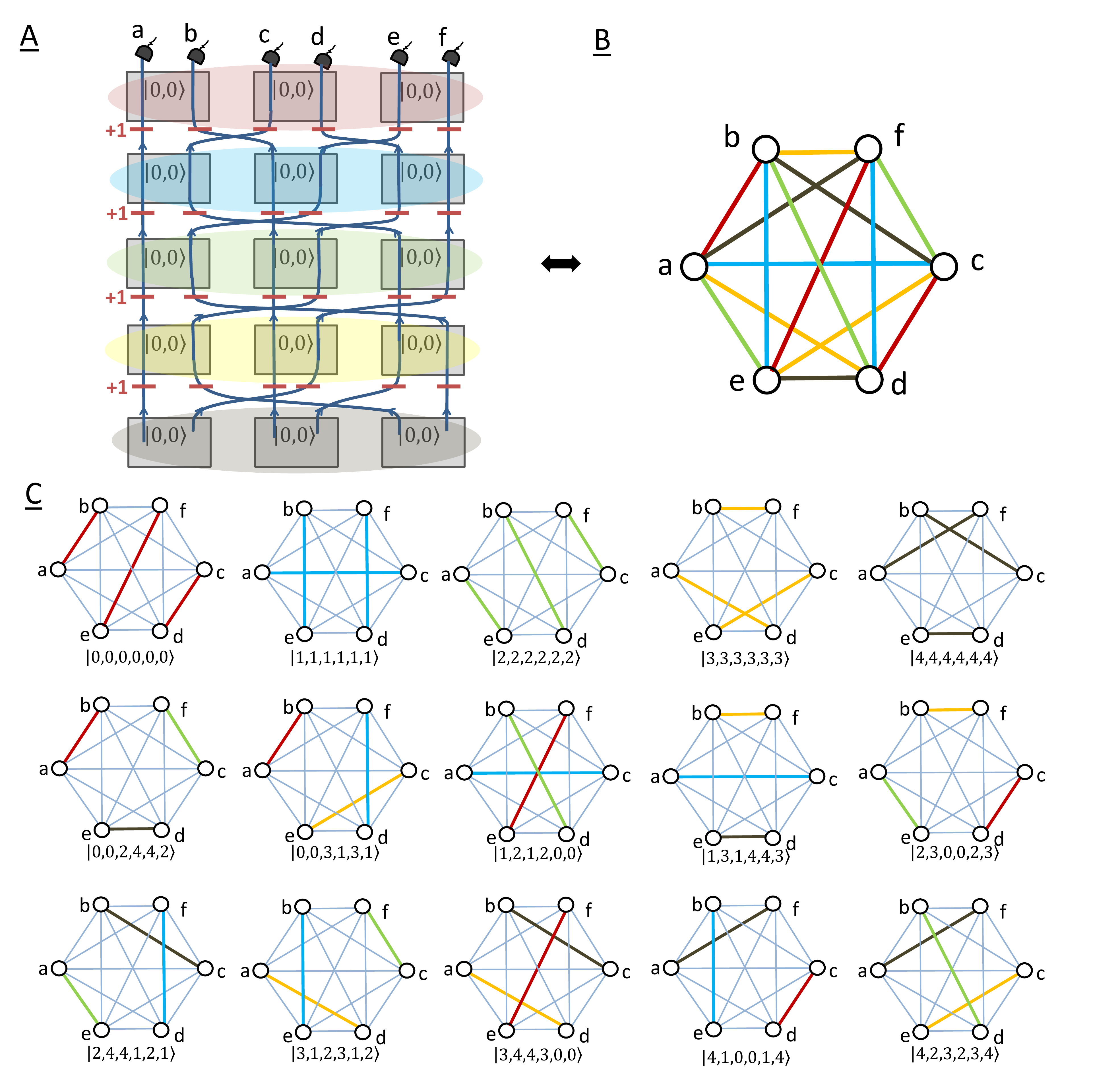}
\caption{\textbf{A}: An experimental setup with five layers with three crystals each, which creates a 6-photon entangled state in five dimensions. \textbf{B}: It is represented by the complete graph $K_6$, and each of the five layers corresponds to one perfect matching (indicated by the edges with the same colors). \textbf{C}: A complete graph with six vertices has 15 perfect matchings. Five of them (first line) correspond to the five different layers which can be arbitrarily controlled in the experiment. The remaining ten perfect matchings (second and third line) correspond to combinations from different layers of crystals.}  
\label{figSI:sixphoton5dim}
\end{figure*}

An experiment with five layers and three crystals in each layer is shown in Fig. \ref{figSI:sixphoton5dim}A. It corresponds to the complete graph $K_6$, which has one edge between each of its six vertices Fig. \ref{figSI:sixphoton5dim}B. It has 15 perfect matchings, which are shown in Fig. \ref{figSI:sixphoton5dim}C. For complete graphs $K_{2n}$ with $2n$ vertices, the number of perfect matchings is $\#(PM)=\frac{(2n)!}{n!2^n}$ \cite{oeisA001147}.

A 1-Factorization of the graph G(V,E) is a partitioning of the graph's edges into disjoined subgraphs (called 1-factors), such as the first line of Fig. \ref{figSI:sixphoton5dim}C. Each of these 1-Factor of the 1-Factorization can be controlled independently in the quantum experiment, the additional Maverick terms arise then automatically. Interestingly, in contrast to factorization of natural numbers, 1-Factorizations of graphs are not unique \cite{oeisA000438}. This allows for a lot of extra variability in the generation of entangled states.

\newpage

\subsection{Appendix V. Bipartite Graphs}
\begin{figure*}[!h]
\includegraphics[width=0.67 \textwidth]{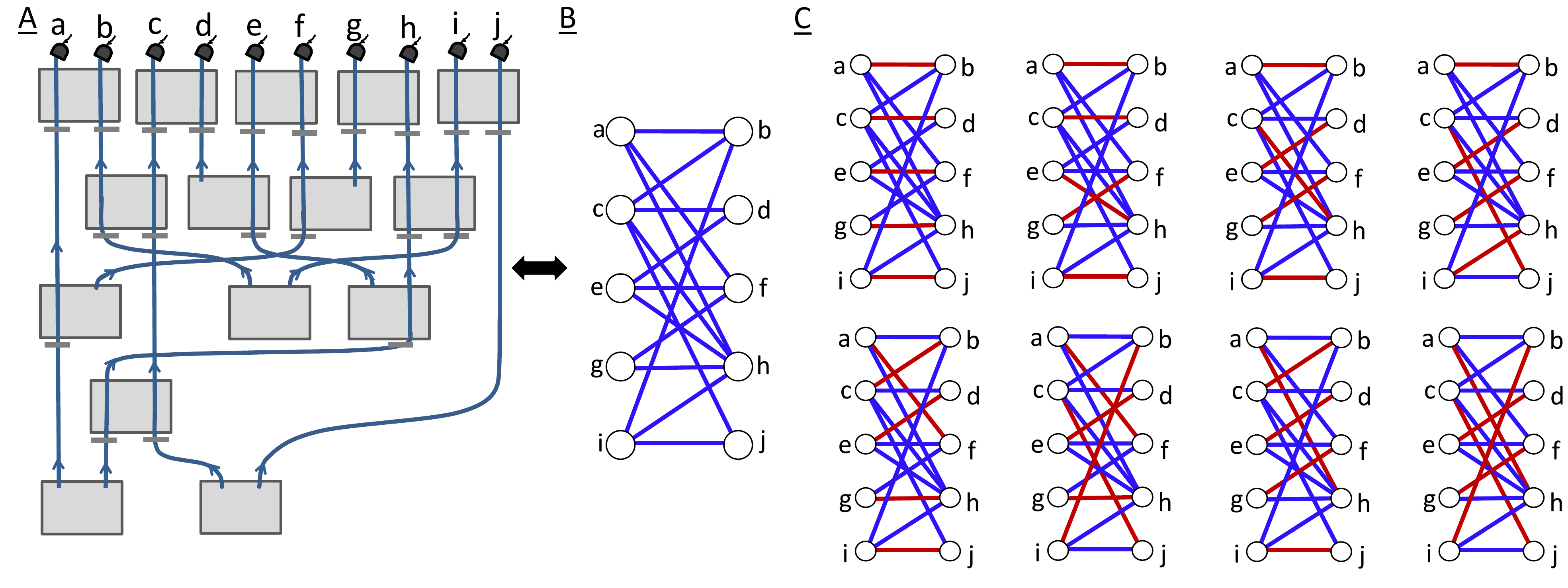}
\caption{\textbf{A}: An optical setup which corresponds to a bipartite graph. It has ten paths and 15 crystals. \textbf{B}: The corresponding bipartite graph. The question how many terms the resulting quantum state will have is asking how many perfect matchings there are in the bipartite graph. \textbf{C}: In this example, there are eight perfect matchings, which are represented with red coloured edges.}  
\label{figSI:Bipartite}
\end{figure*}

Counting the number of perfect matchings in a bipartite graph is in the complexity class \textsc{\#P-complete}. In Fig. \ref{figSI:Bipartite}A, an experimental setup is shown which corresponds to the bipartite graph in Fig. \ref{figSI:Bipartite}B. The perfect matchings for this case can be found in Fig. \ref{figSI:Bipartite}C. They correspond to the number of terms in the resulting quantum state. The mode number of the different terms can be set for each crystal individually, thus one can simply see which states are possible. 

\subsection{Appendix VI. Perfect matchings in general Graphs: Tutte's theorem}
\begin{figure}[h]
\includegraphics[width=0.45 \textwidth]{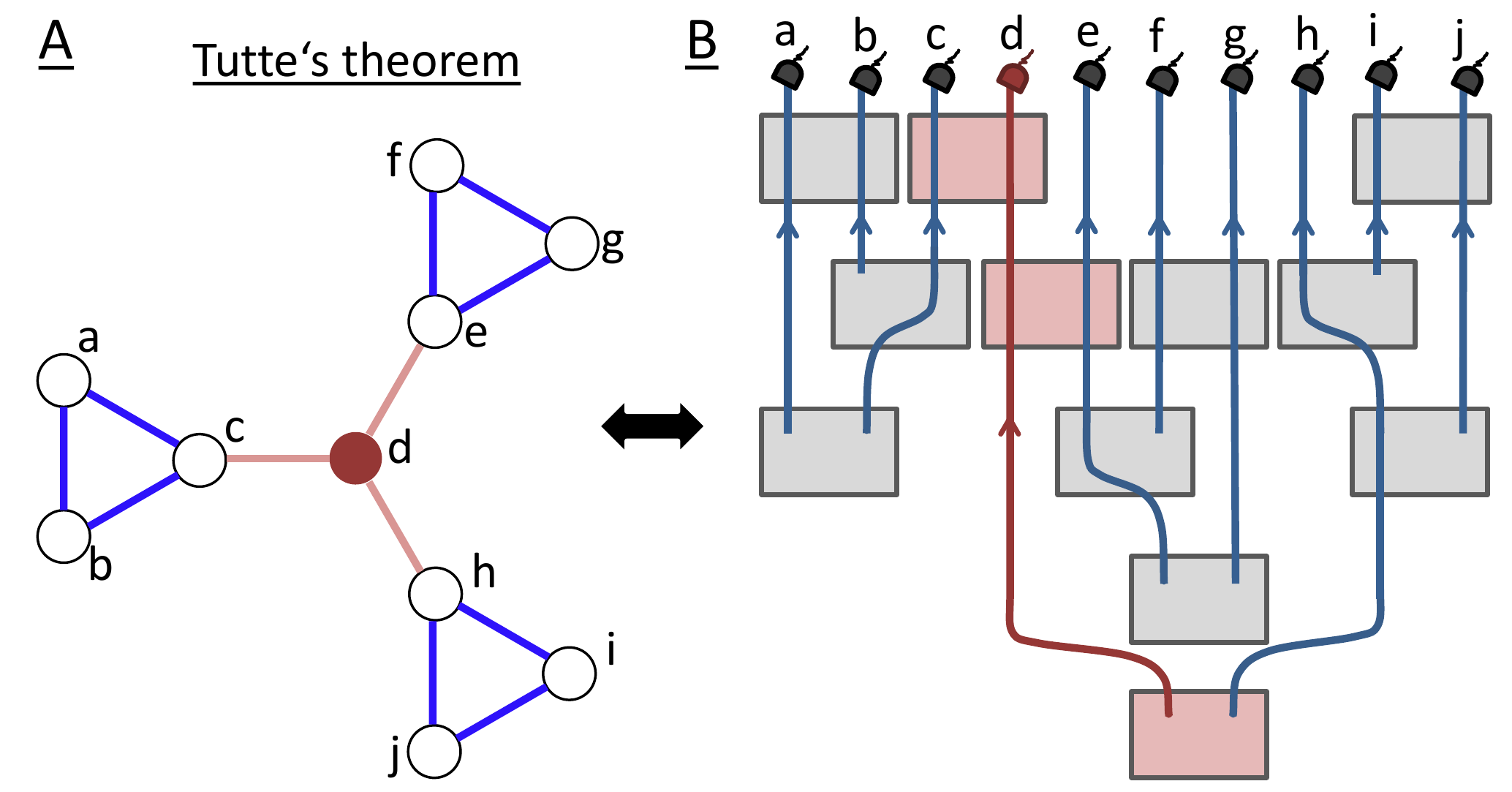}
\caption{\textbf{A}: Tutte's theorem is a generalisation for arbitrary graphs. It says that in a graph $G(V,E)$ a perfect matching exists if and only if for every subset $U \in V$, the remaining subgraph $V-U$ has at most $U$ connected components with an odd number of vertices. In the above example, if we chose $U=d$, the remaining subgraph has three connected components (abc, efg, hij), and each of them has an odd number of vertices. $U$ has only one vertex, thus there is no perfect matching in this graph. \textbf{B}: The analog criterion for a general setup where each crystal produces indistinguishable photon pairs can be states as follows: For every combination of paths $U$, removing the paths and all connected crystals leads to several independent remaining setups $S_r$. Coincident counts can only occur if the number of $S_r$ with odd numbers of paths is smaller than the number of paths in $U$. In the example, the subset $U=d$ does not fulfill the condition: By removing the path $d$ and every connected crystal (depicted in red), $S_r$ contains three independent subsetups (with paths abc, efg, hij), each of them have an odd number (three) of paths. It can be easily understood that subsetups with an odd number of crystals require one photon from the removed subset. If the number of subsetups, which require one photon, is larger than the number of paths removed, not every subsetup will receive a photon, thus there can not be an 2n-fold coincidence count.}  
\label{figSI:tutte}
\end{figure}

A different important result in Graph theory about perfect matchings is \textit{Tutte's theorem}. It gives a necessary and sufficient condition for general graphs, when one can find perfect matchings (but not talking about how many). It is a generalisation of Hall's marriage theorem, which answers the same question for bipartite graphs. In Fig. \ref{figSI:tutte}A, the theorem is explained based on an example. That theorem can be understood with quantum experiments, as shown in Fig. \ref{figSI:tutte}B.
 
\newpage

\subsection{Appendix VII. From Graph to Experimental Setup}
We show one simple example how to construct the experimental setup and the wiring of the paths for one specific graph. We use the random graph shown in Fig3 in the main text.
 
\begin{figure}[h]
\includegraphics[width=1 \textwidth]{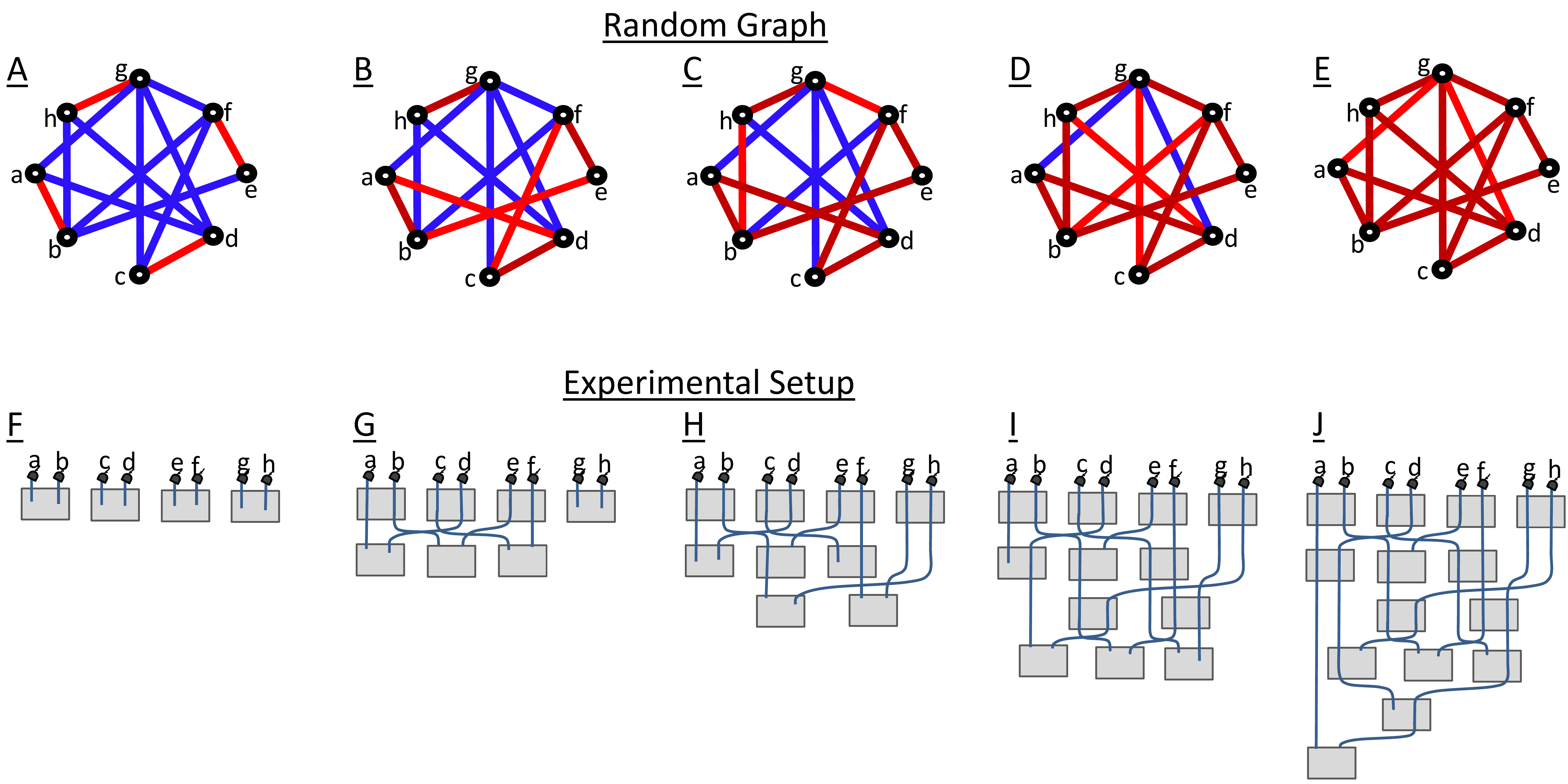}
\caption{\textbf{A}: Generation of an experimental setup from a random graph: \textbf{A-E}: Random graph with increasingly number of edges translated to an experiment. Blue edges correspond to crystals which need to be added to the setup, red edges are those that already correspond to a crystal in the setup. \textbf{F-J}: Experimental setup is constructed according to the graph. First, detectors are set, and setups are the first row of crystal is added. Afterwards, the remaining layers of crystals are added.}  
\label{figSI:construct}
\end{figure}

\end{document}